\documentclass[sigconf, nonacm]{acmart}

\newcommand\vldbdoi{XX.XX/XXX.XX}
\newcommand\vldbpages{XXX-XXX}
\newcommand\vldbvolume{18}
\newcommand\vldbissue{1}
\newcommand\vldbyear{2025}
\newcommand\vldbauthors{\authors}
\newcommand\vldbtitle{\shorttitle} 
\newcommand\vldbavailabilityurl{URL_TO_YOUR_ARTIFACTS}
\newcommand\vldbpagestyle{plain}
\usepackage{booktabs,multirow}
\usepackage{array}

\begin{document}
\title{Substitutability-Based Graph Node Pricing}

\author{Wang Huiju$^\dagger$} 
\affiliation{%
  \institution{Zhongnan University of Economics and Law}
  \city{Hubei}
  \state{China}
}
\email{wanghj@zuel.edu.cn}

\author{Gao Yuanyuan$^\dagger$} 
\affiliation{%
  \institution{Zhongnan University of Economics and Law}
  \city{Hubei}
  \state{China}
}
\email{gaoyuanyuanxgz@163.com}

\author{Zhengkui Wang}
\affiliation{%
  \institution{Singapore Institute of Technology}
  \city{}
  \state{Singapore}
}
\email{Zhengkui.wang@singaporetech.edu.sg}

\author{Yue Xiao$^*$}
\affiliation{%
  \institution{Zhongnan University of Economics and Law}
  \city{Hubei}
  \state{China}
}
\email{yuexiao@zuel.edu.cn}

\begin{abstract}
In the era of data commodification, the pricing of graph data presents unique challenges that differ significantly from traditional data markets. This paper addresses the critical issue of node pricing within graph structures, an area that has been largely overlooked in existing literature. We introduce a novel pricing mechanism based on the concept of substitutability, inspired by economic principles, to better reflect the intrinsic value of nodes in a graph. Unlike previous studies that assumed known prices for nodes or subgraphs, our approach emphasizes the structural significance of nodes by employing a dominator tree, utilizing the Lengauer-Tarjan algorithm to extract dominance relationships. This innovative framework allows us to derive a more realistic pricing strategy that accounts for the unique connectivity and roles of nodes within their respective networks. Our comparative experiments demonstrate that the proposed method significantly outperforms existing pricing strategies, yielding high-quality solutions across various datasets. This research aims to contribute to the existing literature by addressing an important gap and providing insights that may assist in the more effective valuation of graph data, potentially supporting improved decision-making in data-driven environments.
\end{abstract}

\maketitle

\pagestyle{\vldbpagestyle}
\begingroup\small\noindent\raggedright\textbf{PVLDB Reference Format:}\\
\vldbauthors. \vldbtitle. PVLDB, \vldbvolume(\vldbissue): \vldbpages, \vldbyear.\\
\href{https://doi.org/\vldbdoi}{doi:\vldbdoi}
\endgroup
\begingroup
\renewcommand
\thefootnote{}
\footnotetext{$^\dagger$ These authors contributed equally.}
\footnotetext{$^*$ Corresponding author.}
\\
\footnote{\noindent
This work is licensed under the Creative Commons BY-NC-ND 4.0 International License. Visit \url{https://creativecommons.org/licenses/by-nc-nd/4.0/} to view a copy of this license. For any use beyond those covered by this license, obtain permission by emailing \href{mailto:info@vldb.org}{info@vldb.org}. Copyright is held by the owner/author(s). Publication rights licensed to the VLDB Endowment. \\
\raggedright Proceedings of the VLDB Endowment, Vol. \vldbvolume, No. \vldbissue\ %
ISSN 2150-8097. \\
\href{https://doi.org/\vldbdoi}{doi:\vldbdoi} \\
}\addtocounter{footnote}{-1}\endgroup

\ifdefempty{\vldbavailabilityurl}{}{
\vspace{.3cm}
\begingroup\small\noindent\raggedright\textbf{PVLDB Artifact Availability:}\\
The source code, data, and/or other artifacts have been made available at \url{\vldbavailabilityurl}.
\endgroup
}

\section{INTRODUCTION}

Graphs have emerged as a powerful modeling tool, widely utilized to represent networks across various domains ~\cite{majeed2020graph, Tang2010, bti1105, 978-3-319-64468-45, 2015Efficient}, such as social networks ~\cite{social}, citation networks ~\cite{cite}, and communication networks ~\cite{commun}. These applications highlight the critical role of graph structures in capturing interconnected entities and their interactions. Currently, the global data economy has experienced exponential growth, driven by the commodification of data ~\cite{analytics2016age}, platforms like Shanghai Data Exchange and Guangzhou Data Exchange have become key players in the data marketplace and several prominent data trading platforms, such as DataMarket, Kaggle and Amazon Web Services have witnessed a surge in trading volume.

The rapid expansion of data trading platforms hinges crucially on effective data pricing mechanisms, with graph node valuation emerging as a critical frontier. As a core algorithmic component in data marketplaces, node pricing directly influences transaction efficiency and market liquidity. Unlike homogeneous datasets sold on platforms like DataMarket, graph nodes derive economic value from their topological positions, connection strengths, and functional roles within networked systems. For instance, a node representing a key influencer in a social network graph commands higher value due to its ability to propagate information, while a node in a transportation network might be priced based on its centrality in optimizing logistics routes. Or in supply chain networks, a single strategically positioned warehouse node controlling multiple delivery routes can enable cost-efficient operations, while an overlooked supplier node could trigger cascading failures in production lines. However, despite its practical relevance, graph node pricing remains underexplored compared to structured data. Unlike tabular data, graph nodes derive value from their topological roles and interdependencies, rendering traditional pricing models inadequate. Existing approaches often oversimplify node independence or ignore structural nuances, leading to suboptimal valuations.

Previous studies on graph data pricing have primarily focused on influence propagation model and query-driven framework. For example, Sun et al. ~\cite{SUN2024122815} proposed a graph autoencoder-based method for social network node pricing, leveraging predicted influence and approximate Shapley values to align prices with marketing value. Query-centric methods, such as GQP proposed an arbitrage-free framework reusing precomputed price points for dynamic queries ~\cite{chen2022gqp}. Scenario-specific methods further address incomplete graph scenarios: Hou et al. designed a discount pricing function to handle missing information ~\cite{hou2023scalable}. Such approaches face two key limitations: (1) Its reliance on initial cascade graphs for influence prediction limits applicability in scenarios lacking historical propagation data. (2) The availability of known node or subgraph prices, and the equality of node prices across the graph. However, these premises or assumptions are not entirely applicable in the real data trading environment. On the one hand, it is difficult to obtain historical orders for current data transactions, and some data that has not been traded also needs to be priced reasonably; On the other hand, these assumptions oversimplify the intrinsic complexity of node pricing in real-world applications.

Existing methods for measuring node importance in graph theory, such as degree centrality and eigenvector centrality, exhibit significant limitations in accurately assessing node value. These methods often treat edges as homogeneous entities, disregarding their unique roles and structural significance within the network. For instance, degree centrality focuses solely on the number of connections a node has, while eigenvector centrality considers the influence of a node’s neighbors. Although these methods do not require historical data for training, these approaches primarily emphasize local characteristics or global structural properties without adequately capturing the nuanced interdependencies and contextual relevance of nodes within the graph. Moreover, traditional pricing methodologies based on revenue allocation, such as those leveraging the Shapley value, typically assume that nodes are independent and are better suited for structured data ~\cite{fatima2008linear, chen2023algorithms}. However, this assumption does not hold for graph data due to the inherent structural dependencies among nodes ~\cite{himelboim2017classifying}. In graph structures, nodes are interconnected through edges, and their value is significantly influenced by their relationships and positions within the network. The independence assumption oversimplifies the complexity of real-world graph structures, leading to inaccurate pricing outcomes that fail to capture the nuanced interdependencies and contextual relevance of nodes. Thus, node pricing constitutes a foundational challenge for graph data markets, necessitating methods that explicitly model structural significance and substitutability.

To address these gaps, we introduce substitutability: a concept rooted in economic theory ~\cite{reca2006issues}, as a novel lens for node valuation. In economics, substitutability quantifies how easily one product can replace another while maintaining utility ~\cite{sc}. Translating this to graphs, nodes with overlapping structural roles (e.g., similar connectivity patterns) exhibit higher substitutability, reducing their individual value. Conversely, nodes occupying unique positions (e.g., dominators in shortest paths) are irreplaceable, warranting premium pricing. Consider a transportation network: A hub city through which all east-west rail lines pass cannot be substituted without disrupting connectivity, whereas towns with redundant routes may compete on price. Substitutability thus provides a dual perspective: It balances local similarity (via path overlap) and global criticality (via dominance hierarchies) to reflect real-world pricing. Existing utility-based approaches such as destructiveness and influence analysis fail to microscopically evaluate node substitutability from the perspective of data transactions. Specifically, destructiveness metrics quantify a node's impact on the entire graph through macroscopic analysis. For example, two nodes with identical destructiveness scores of 50 may not be mutually substitutable, as their functional values could diverge significantly due to topological roles and economic attributes. This discrepancy highlights inherent limitations in traditional methods, which overlook the micro-level substitutability relationships critical for data trading scenarios. Hence, there is an urgent need to develop substitutability evaluation frameworks that explicitly account for node-specific functional values and transactional contexts.

Our approach operationalizes substitutability through dominator tree, which hierarchically model node indispensability. The dominator tree, constructed via the Lengauer-Tarjan algorithm [11], identifies nodes that act as mandatory gatekeepers for others. For example, in a supply chain network, a warehouse dominating all paths to regional distributors would exhibit low substitutability. To quantify substitutability, we integrate two components: (1) Positional criticality, derived from dominator trees, measures how many nodes depend on a target node for connectivity; (2) Path similarity, computed via overlapping incoming/outgoing paths, evaluates how structurally interchangeable two nodes are. By synthesizing these metrics, our method assigns higher prices to nodes that are both topologically critical and structurally unique. For instance, in a social network, a user bridging disparate communities (high criticality) with no overlapping influence regions (low path similarity) would command a higher price than a user with redundant connections.  

This paper proposes a method for pricing nodes in a graph based on substitutability, utilizing the dominator tree concept to extract the dominance relationships of nodes. This innovative approach allows for a more realistic pricing mechanism that reflects the structural dynamics of graph data. Thus, we make the following contributions to tackle the node pricing problem:

\begin{itemize}
\item We are the first work focusing on the node pricing problem for graph data in general scenarios, diverging from previous studies that assumed node or subgraph prices were known.
\item We are the first to propose the concept of graph node substitutability, providing a more realistic mechanism for node valuation.
\item We employ the dominator tree algorithm alongside the Lengauer-Tarjan method ~\cite{dtree} and path similarity to consider structural information, allowing us to elucidate substitutability in node pricing.
\item Comparative experiments demonstrate that our node pricing method yields high-quality solutions and significantly outperforms existing methods.
\end{itemize}

The rest of the paper is organized as follows. Section 2 introduces the problem definition. Section 3 presents node pricing method. Section 4 describes the experiments. Section 5 reviews the related work.  Finally, we conclude the paper in Section 6.

\section{PROBLEM DEFINITION}

In this section, we define basic concepts for node pricing in graph data. \autoref{tab:1} summarizes the abbreviations and notations used in this paper.

\begin{table}[hb]
  \caption{Abbreviation and Notation}
  \label{tab:1}
  \begin{tabular}{ccl}
    \toprule
    Symbol & Description\\
    \midrule
    \textit{G} & Graph data\\
    \textit{$S(v_i, v_j)$} & Node similarity\\
    \textit{$B(v_i)$} & Node substitutability\\
    \textit{$c(v_i)$} & Node position coefficient\\
    \textit{$e(v_i)$} & The path that can reach the current node\\
    \textit{$r(v_i)$} & The path starting from the current node\\
    \textit{$p$} & Node price\\
  \bottomrule
\end{tabular}
\end{table}

\textbf{Definition 1: Graph data commodity.} Given a directed graph \textit{G = (V, E)}, where (1) \textit{V} is a finite set of nodes; (2) \textit{E} $\subseteq$ \textit{V} $\times$ \textit{V} is a set of directed edges. 

Graph nodes are treated as homogeneous commodities in this paper. Unlike conventional goods, node data products differ primarily in their information content, while exhibiting no significant differences in other aspects. This characteristic aligns them with homogeneous goods in economics, leading us to treat node products as homogeneous for the purposes of this discussion. Due to this characteristic, homogeneous goods generally possess substitutability \cite{baijmol1972economic}. 

\textbf{Definition 2: Node substitutability.} The substitutability $B$ of node quantifies its replaceability by other nodes in fulfilling equivalent structural roles. The node substitutability is defined as:

\begin{equation}
B(v_i)=\frac{1}{\vert V \vert -1}\sum_{j\neq i}(c(v_i)S(v_i, v_j))
\end{equation}

where $c(v_i)$ quantifies $v_i$'s positional criticality in maintaining graph connectivity, and $S(v_i, v_j)$ measures path similarity between $v_i$ and $v_j$. 

\textbf{Definition 3: Path Similarity.} The structural similarity $S$($v_i$, $v_j$) between node $v_i$ and $v_j$ is defined as the overlap of their incoming paths $e$($\cdot$) and outgoing paths $r$($\cdot$):

\begin{equation}
S(v_i, v_j) = Average(\frac{\vert r(v_i) \cap r(v_j) \vert}{r(v_i)}, \frac{\vert e(v_i) \cap e(v_j) \vert}{e(v_i) } )
\end{equation}

\begin{figure}
  \centering
  \includegraphics[width=0.8\linewidth]{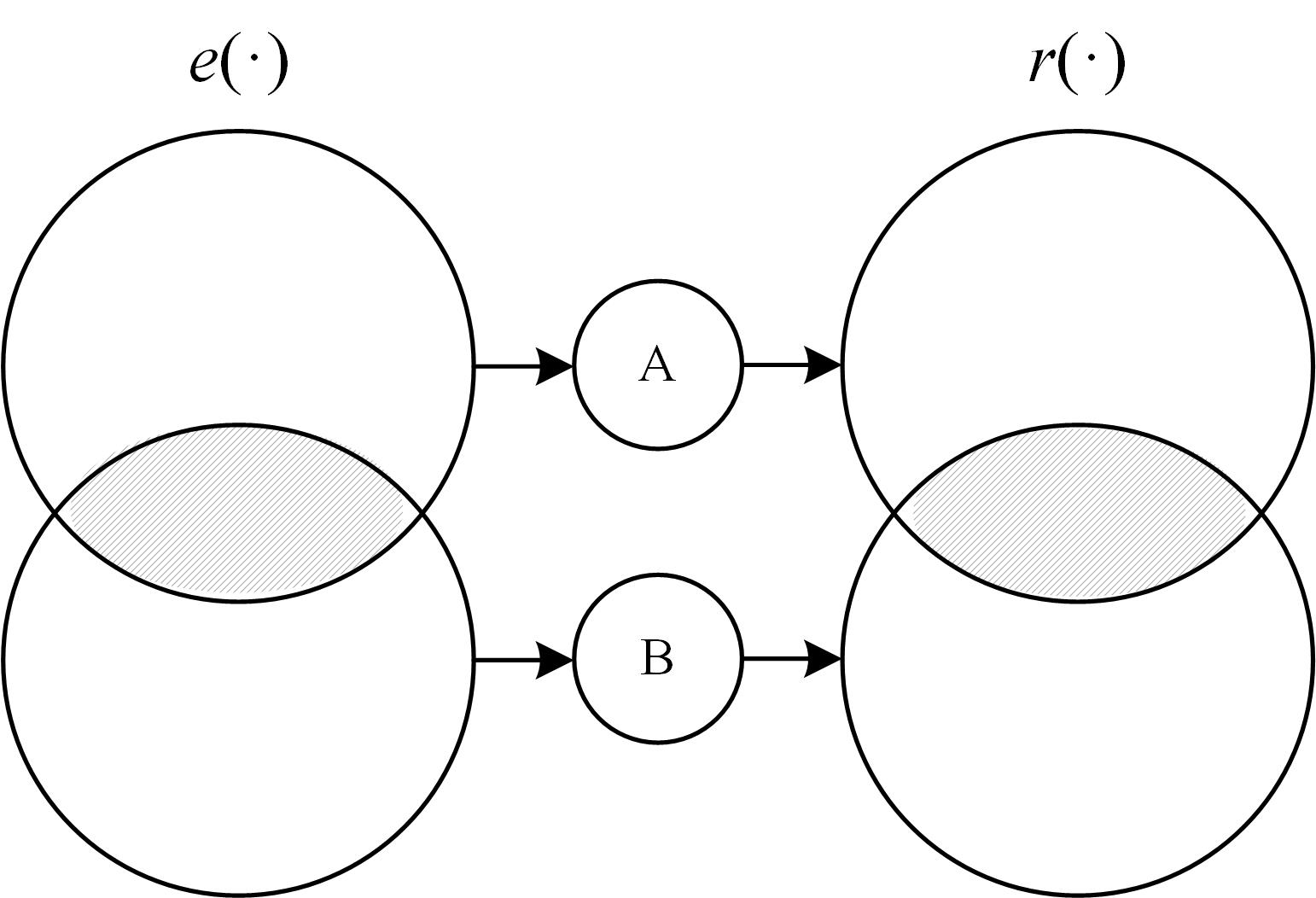}
  \caption{Example of Node Path Similarity.}
  \label{fig:sim}
\end{figure}

As illustrated in \autoref{fig:sim}, the intersection of paths between nodes A and B represents the portion of similarity, while the remaining paths indicate the irreducible aspects of the other node.

\textbf{Definition 4: Node positional criticality.} The positional criticality $c(v_i)$ of node $v_i$ quantifies its indispensability in maintaining the connectivity and structural integrity of the graph. Simply, the positional criticality $c(v_i)$, which measures its dominance over other nodes by quantifying how many nodes depend on $v_i$ to maintain connectivity, can be defined as:

\begin{equation}
c_i = \frac{Number\_of\_nodes\_requiring\_v_i\_in\_their\_shortest\_paths}{Total\_nodes - 1}
\end{equation}

where its dependency can be described that A node $v_j$ "requires" $v_i$ if all shortest paths from a designated root node (e.g., the graph’s entry point) to v$_j$ pass through $v_i$.

\textbf{Definition 5: Node pricing.} We price nodes based on the relationship between substitutability and commodity price as established in economics. In this context, consider two substitutable entities, where $Q_i$ represents the quantity (demand) of the commodity. Typically, the price $P_i$ and and the substitutability $b$ satisfy the following linear demand function \cite{singh1984price}:

\begin{equation}
P_i = A_i-a_iQ_i-bQ_i, (i=1, 2, i\neq j)
\end{equation}

\begin{equation}
Q_i = \frac{a_iA_i-bA_i}{a_ia_j-b^2}-\frac{a_j}{a_ia_j-b^2}P_i+\frac{b}{a_ia_j-b^2}P_j
\end{equation}

where $A_i$ and $a_i$ are constants. To ensure that the quantities remain positive, it is necessary that the inequality $a_iA_i-bA_i$ > 0.

It can be observed that for substitutable commodities, the stronger the substitutability (the larger the value of $b$), the lower the price $P_i$.

In this paper, we treat nodes as a form of data commodity, and we can draw parallels to the economic principle that greater substitutability leads to lower prices, thus, node price is defined as:

\begin{equation}
p=\frac{log(-B(v_i))}{\sum_{i=1}^nlog(-B(v_i))}
\end{equation}

\begin{figure}
  \centering
  \includegraphics[width=0.8\linewidth]{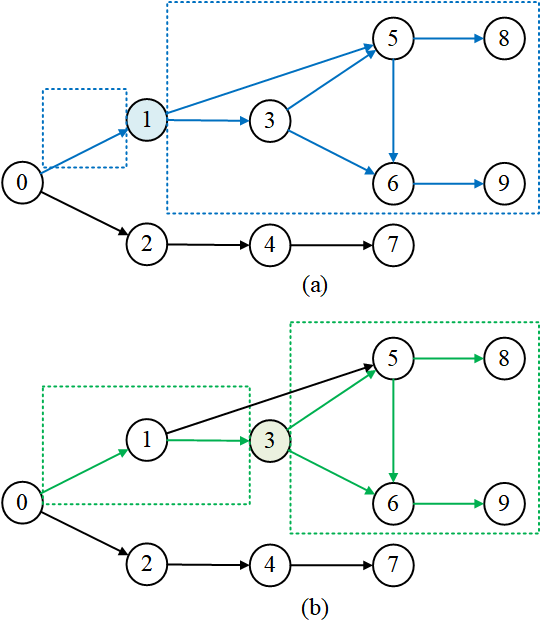}
  \caption{Example of Node Substitutability.}
  \label{fig:path}
\end{figure}

\textbf{Example 1.} We provide an example to illustrate the rationale behind calculating node substitutability in this manner. As illustrated in the directed graph shown in \autoref{fig:path}, we take nodes 1 and 3 as examples. Assuming we determine the substitutability of nodes 1 and 3 based on the nodes present in their respective paths, it may appear that they possess consistent structural information, suggesting that the two nodes can be fully interchangeable. However, this conclusion is evidently flawed given the structural relationships depicted in the graph. The distinguishing structural information for these two nodes lies in the differences in the paths traversing them, including their in-degrees and out-degrees. Therefore, we proceed to calculate their substitutability based on path similarity as the guiding principle.

By performing calculations, we can derive the sets of paths traversing these two nodes (\autoref{tab:path}). Based on this information, we can further determine the intersection of the two sets of paths. This intersection provides valuable insight into the structural relationships and similarities between the nodes, serving as a foundation for assessing their substitutability within the graph.

\begin{table}[hb]
  \caption{Node Path Set}
  \label{tab:path}
  \newcolumntype{P}[1]{>{\raggedright\arraybackslash}p{#1}}
  \begin{tabular}{p{0.5cm}p{3.6cm}p{1cm}p{1.9cm}}
    \toprule
    Node &$e$($\cdot$) &$r$($\cdot$) &Intersection\\
    \midrule
    \textit{1} & \{1 $\rightarrow$ 3, 1 $\rightarrow$ 5, 3 $\rightarrow$ 5, 3 $\rightarrow$ 6, 5 $\rightarrow$ 6, 5 $\rightarrow$ 8, 6 $\rightarrow$ 9\} &\{0 $\rightarrow$ 1\} & \multirow{2}{=}{\{3 $\rightarrow$ 5, 3 $\rightarrow$ 6, 5 $\rightarrow$ 6, 5 $\rightarrow$ 8, 6 $\rightarrow$ 9\} \\ \{0 $\rightarrow$ 1\}} \\
    \textit{3} & \{3 $\rightarrow$ 5, 3 $\rightarrow$ 6, 5 $\rightarrow$ 6, 5 $\rightarrow$ 8, 6 $\rightarrow$ 9\} & \{0 $\rightarrow$ 1, 1 $\rightarrow$ 3\} & \\
    
  \bottomrule
\end{tabular}
\end{table}

Based on the analysis of the path intersection between the two nodes, it can be observed that Nodes 1 and 3 exhibit a high degree of similarity. From a similarity perspective, it can be concluded that Nodes 1 and 3 are substitutable, and the degree of substitutability is considerable.

However, when assessing the substitutability of nodes, relying solely on path similarity is insufficient for accurately evaluating a node's contribution to the entire graph. The position of the nodes is equally critical. For instance, in Example 1, although Nodes 1 and 3 demonstrate a high level of path similarity, this single metric cannot definitively determine the substitutability of the nodes. Specifically, an analysis of the paths leading to the current nodes reveals a parent-child relationship between Nodes 1 and 3. The presence of Node 1 directly influences whether the starting point can reach Node 3. Furthermore, examining the paths originating from the current node, it is evident that while Node 1 has only one additional reachable path, \{1 $\rightarrow$ 5\}, beyond those connected to Node 3, the number of nodes reachable from Node 1 remains unaffected by the existence of Node 3. In other words, the position of Node 1 is more critical compared to Node 3.

Therefore, when discussing the substitutability of nodes, it is essential to not only rely on path similarity, but also to consider the nodes' positions within the entire graph and their interconnections. Consequently, when calculating node substitutability, it is necessary to introduce a positional coefficient.

\section{GRAPH NODE PRICING ALGORITHM AND OPTIMIZATION}

In this section, we propose a garph node pricing algorithm, and employing an approximation algorithm to efficiently compute node substitutability while addressing computational complexity challenges.

\subsection{Basic Graph Data Pricing Algorithm}
The core objective of the basic algorithm is to compute node substitutability $B(v_i)$ by integrating path similarity $S(v_i, v_j)$ and positional criticality $c(v_i)$, as defined in Section 2. The algorithm operates in four stages: (1) path collection, (2) positional criticality calculation, (3) substitutability computation, and (4) price derivation. Algorithm 1 outlines the baseline approach.

\begin{table}[hb]
  \label{tab:a1}
  \begin{tabular}{l}
    \toprule
    \textbf{Algorithm 1:} Basic Graph Data Pricing Algorithm\\
    \midrule
    \textbf{Input:} directed graph \( G = (V, E) \), root node \( r \) \\
    \textbf{Output:} node prices \( p \) for all \( v \in V \) \\
    \textbf{1} \textbf{Path Collection:} \\
    \textbf{\quad}For each node \( v_i \in V \):\\
    \textbf{2 \quad} Compute \( e(v_i) \leftarrow \) set of all incoming paths from \( r \) to \( v_i \) \\
    \textbf{3 \quad} Compute \( r(v_i) \leftarrow \) set of all outgoing paths from \( v_i \)\\
    \textbf{4} \textbf{Positional Criticality Calculation:} \\
    \textbf{\quad} For each node \( v_i \in V \): \\
    \textbf{5 \quad} Initialize \( \text{count} \leftarrow 0 \)\\
    \textbf{6 \quad} For each node \( v_j \in V \setminus \{v_i\} \): \\
    \textbf{7 \qquad} \textbf{if} all shortest paths from \( r \) to \( v_j \) pass through \( v_i \): \\
    \textbf{8 \qquad \quad} \( \text{count} \leftarrow \text{count} + 1 \) \\
    \textbf{9 \quad} \( c(v_i) \leftarrow \frac{\text{count}}{|V| - 1} \) \\
    \textbf{10} \textbf{Substitutability Computation:} \\
    \textbf{\quad} For each node \( v_i \in V \): \\
    \textbf{11 \quad} \( B(v_i) \leftarrow 0 \) \\
    \textbf{12 \quad} For each node \( v_j \in V \setminus \{v_i\} \): \\
    \textbf{13 \qquad \quad} Compute \( S(v_i, v_j) \): \\
    \textbf{14 \qquad \qquad} \( \text{in\_ratio} \leftarrow \frac{|e(v_i) \cap e(v_j)|}{|e(v_i)|} \) \\
    \textbf{15 \qquad \qquad} \( \text{out\_ratio} \leftarrow \frac{|r(v_i) \cap r(v_j)|}{|r(v_i)|} \) \\
    \textbf{16 \qquad \qquad} \( S(v_i, v_j) \leftarrow \frac{\text{in\_ratio} + \text{out\_ratio}}{2} \) \\
    \textbf{17 \qquad \quad} \( B(v_i) \leftarrow B(v_i) + c(v_i) \cdot S(v_i, v_j) \) \\
    \textbf{18 \quad} \( B(v_i) \leftarrow \frac{B(v_i)}{|V| - 1} \)\\
    \textbf{19} \textbf{Price Derivation:} \\
    \textbf{\quad} Compute \( \text{total} \leftarrow \sum_{v_i \in V} \log(-B(v_i)) \) \\
    \textbf{20 }For each node \( v_i \in V \): \\
    \textbf{21 \quad}\( p(v_i) \leftarrow \frac{\log(-B(v_i))}{\text{total}} \) \\
    \textbf{22 }Return \( p \) \\
    
    \bottomrule
\end{tabular}
\end{table}

The algorithm begins with path collection, for each node $v_i$ incoming paths $e(v_i)$ (from root $r$) and outgoing paths $r(v_i)$ are collected via BFS/DFS traversals (line 2-3), and the criticality $c(v_i)$ measures how many nodes $v_j$ depend on $v_i$ in their shortest paths from $r$. This requires checking all pairs of nodes, which is computationally intensive (line 5-9). Next, pairwise path similarity $S(v_i, v_j)$ is computed using Jaccard-like ratios of overlapping paths. Substitutability $B(v_i)$ aggregates these similarities weighted by $c(v_i)$ (line 11-18). Finally, prices are normalized using the logarithmic scaling defined in Equation 6 (line 19-22).

Time Complexity of the algorithm is $O(n^24^n)$, rendering the baseline method infeasible for large graphs. This high complexity underscores the necessity of structural optimizations. In Section 3.2, we introduce the dominator tree to replace brute-force shortest path checks, reducing positional criticality computation to $O(mlogn)$ via the Lengauer-Tarjan algorithm. This optimization drastically improves scalability while preserving accuracy.

\subsection{Dominator Tree-based Algorithm}

The key idea behind our pricing mechanism is rooted in the concept of substitutability. In calculating the substitutability of nodes, we initially utilized path similarity. However, our analysis revealed that path similarity alone does not fully capture a node's "critical position" within the entire graph. To address the limitations of path similarity in quantifying structural criticality, we introduce dominator tree as a complementary mechanism to capture essential control dependencies within the graph. While path similarity effectively measures node interchangeability through overlapping path patterns, it fails to account for hierarchical control relationships where certain nodes act as mandatory gatekeepers for others. Dominator trees explicitly model these indispensable relationships by revealing which nodes "must" be traversed to reach specific regions of the graph. This structural perspective complements substitutability analysis by identifying nodes that cannot be bypassed – nodes with high dominance exert inherent influence over large subgraphs, making them structurally irreplaceable even if path similarity metrics suggest otherwise. 

Given a graph $G$, considering a node $d$ in a dominator tree is said to dominate another node $v$ if every path from the entry node to $v$ must pass through $d$. In other words, $d$ is an ancestor of $v$ in the dominator tree. The root of the dominator tree is the entry node of the program and dominates all other nodes. This method can extract the dominance relationships of nodes in a directed graph, so that we can further analyze their structural importance from the perspectives of substitutability. 

In addition, to improve algorithm efficiency, we use the Lengauer-Tarjan Algorithm fusion dominator tree ~\cite{dtree}, and the time complexity of the dominator tree with the Lengauer-Tarjan Algorithm is $O(m logn)$, where $m$, $n$ is the number of edges and nodes in the graph, respectively.

\begin{figure}
  \centering
  \includegraphics[width=\linewidth]{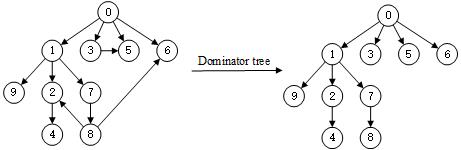}
  \caption{The effect of transforming a directed graph into a dominator tree for graph data.}
  \label{fig:1}
\end{figure}

\textbf{Example 2.} Given a data graph \textit{G}, \autoref{fig:1} demonstrates the effect of transforming a directed graph into a dominator tree for graph data $G_t$, omitting some edges that are irrelevant to the dominating relationship, such as the directed edges from nodes 3 to 5, nodes 8 to 2, and nodes 8 to 6 in the graph.

\subsection{Path Similarity-based Node Substitutability Algorithm.} 

The main functions of our node pricing algorithm are shown in Algorithm 2. The algorithm first aims to efficiently compute the incoming and outgoing edges for each node in the dominator tree, which is crucial for the overall pricing mechanism. By leveraging a Depth-First Search (DFS) approach, we can traverse the tree and collect the necessary edge information.

The algorithm begins by reading the tree structure from file and building a bidirectional graph representation (lines 1-3). Each node maintains sets of incoming edges and outgoing edges. For each target node (line 5), the algorithm first collects all potential incoming edges through backward DFS traversal (line 6) and outgoing edges through forward DFS traversal (line 7). Then it iterates through all other nodes in the graph (lines 8-9), calculating the intersection ratios of both incoming and outgoing edges between the current node and other nodes (lines 10-13). Finally, the algorithm computes the normalized overlap ratio by averaging over all node pairs (line 14) and outputs the results (line 16). From the analysis, we can obtain the algorithm's time complexity is $O$($n^2$).

\begin{table}[hb]
  \label{tab:a2}
  \begin{tabular}{l}
    \toprule
    \textbf{Algorithm 2:} Node Substitutability Algorithm\\
    \midrule
    \textbf{Input:} dominator tree $G_t=(V, E_t)$ \\
    \textbf{Output:} Node similarity ratios for node substitutability\\
    \textbf{1} $tree$ $\leftarrow$ ReadTree (file\_path) \\
    \textbf{2} $graph$ $\leftarrow$ BuildGraph ($tree$)\\
    \textbf{3} $nodes$ $\leftarrow$ GetNodes (graph)\\
    \textbf{4} $overlap\_ratios$ $\leftarrow$ empty dict\\
    \textbf{5 for} each $node$ in $nodes$ do:\\
    \textbf{6 \quad} $incoming$ $\leftarrow$ FindAllIncomingEdges ($graph$, $node$) \\
    \textbf{7 \quad} $outgoing$ $\leftarrow$ FindAllOutgoingEdges ($graph$, $node$) \\
    \textbf{8 \quad} $total\_in$ = len($incoming$) \\
    \textbf{9 \quad} $total\_out$ = len($outgoing$) \\
    \textbf{10\quad}$overlap\_sum$ = 0 \\
    \textbf{11\quad}\textbf{for} $other\_node$ in $nodes$ where $other\_node$ $\neq$ $node$ do:\\
    \textbf{12 \qquad} $other\_in$ = FindAllIncomingEdges ($graph$, $other\_node$)\\
    \textbf{13 \qquad} $other\_out$ = FindAllOutgoingEdges ($graph$, $other\_node$)\\
    \textbf{14 \qquad} $in\_ratio$ = $\vert$ $incoming$ $\cap$ $other\_in$ $\vert$ / $total\_in$ \\
    \qquad \qquad \qquad \qquad if $total\_in$ > 0 else 0 \\
    \textbf{15 \qquad} $out\_ratio$ = $\vert$ $outgoing$ $\cap$ $other\_out$ $\vert$ / $total\_out$  \\
     \qquad \qquad \qquad \qquad if $total\_out$ > 0 else 0 \\
    \textbf{16 \qquad} $overlap\_sum$ += ($in\_ratio$ + $out\_ratio$) / 2\\
    \textbf{17 \quad}$overlap\_avg$ = $overlap\_sum$ / ($\vert$ $nodes$ $\vert$ - 1)\\
    \textbf{18 \quad}$overlap\_avg$[$node$] = $overlap\_avg$\\
    \textbf{19 } OutputToFile ($overlap\_ratios$)\\

    \bottomrule
\end{tabular}
\end{table}

\textbf{Example 3.} Let's provide an example to supplement the above algorithm. \autoref{fig:1} summarizes an example of graph. Through our calculations, we can obtain the similarity for each node, as presented in \autoref{tab:simil}.

\begin{table}[hb]
  \caption{Node similarity}
  \label{tab:simil}
  \begin{tabular}{p{3cm}p{3cm}}
    \toprule
    Node & Node similarity \\
    \midrule
    0 & 0.0432 \\
    1 & 0.3556 \\
    2 & 0.2778 \\
    3 & 0.0000 \\
    4 & 0.1111 \\
    5 & 0.0000 \\
    6 & 0.0000 \\
    7 & 0.2778 \\
    8 & 0.1111 \\
    9 & 0.1389 \\
  \bottomrule
\end{tabular}
\end{table}

Based on the analysis, it is evident that nodes 3, 5, and 6 exhibit dissimilar path characteristics compared to other nodes. This observation can be attributed to the fact that these nodes are directly connected to the root node and do not possess any additional child nodes, resulting in their path sets having no overlap with those of other nodes. Consequently, this calculation suggests that nodes 3, 5, and 6 are irreplaceable and hold the highest value among all nodes. However, such a conclusion is evidently insufficiently rational. To address this issue, we propose assigning an average price to these types of nodes during the pricing calculation. This average price reflects their role in the graph as having a unique path that is irreplaceable by other nodes, while also indicating that this is the only path they possess.

\subsection{Approximation Algorithm to Node Substitutability} 

The structural similarity analysis in large-scale graphs faces fundamental scalability challenges due to its inherent $O(n^2)$ computational complexity when using exact pairwise comparisons. In real-world network datasets (e.g., social networks with >$10^6$ nodes), this quadratic growth renders precise calculations computationally prohibitive and memory-intensive.

The MinHash-LSH approximation addresses these limitations through two key insights: (1) Structural roles can be effectively encoded via edge set fingerprints rather than exhaustive comparisons, and (2) Locality-sensitive hashing enables sublinear-time similarity searches without full pairwise evaluations. The algorithm achieves $O(n)$ time complexity compared to the original $O(n^2)$ exact method, making it suitable for large-scale graphs.

Algorithm 3 shows the details of approximate node similarity calculation using MinHash-LSH. The algorithm begins by constructing the graph structure and converting edge sets into compact MinHash signatures (lines 1-5). For each node's incoming/outgoing edges (line 7-8), it generates probabilistic fingerprints using permutation hashing (lines 9-10). These signatures are indexed in LSH buckets for efficient similarity search (line 12). During query phase (lines 15-18), the algorithm finds candidate nodes with similar connectivity patterns through parallel hash table lookups. The final similarity score reflects the proportion of nodes sharing structural patterns in the graph. 

\begin{table}[hb]
  \label{tab:a2}
  \begin{tabular}{l}
    \toprule
    \textbf{Algorithm 3:} Node Substitutability with MinHash-LSH\\
    \midrule
    \textbf{Input:} dominator tree $G_t=(V, E_t)$, $num\_perm$, $threshold$ \\
    \textbf{Output:} Approximate similarity ratios\\
    \textbf{1} $tree$ $\leftarrow$ ReadTree (file\_path) \\
    \textbf{2} $graph$ $\leftarrow$ BuildGraph ($tree$)\\
    \textbf{3} Initialize LSH index with parameters $num\_perm$, $threshold$\\
    \textbf{4 for} each $node$ in $graph$ do:\\
    \textbf{5 \quad} $in\_edges$ $\leftarrow$ $graph[node].incoming$ \\
    \textbf{6 \quad} $out\_edges$ $\leftarrow$ $graph[node].outgoing$ \\
    \textbf{7 \quad} $m\_in$ $\leftarrow$ CreateMinHash($in\_edges$) \\
    \textbf{8 \quad} $m\_out$ $\leftarrow$ CreateMinHash($out\_edges$) \\
    \textbf{9 \quad} LSH.insert($node$, $m\_in$) \\
    \textbf{10\quad}LSH.insert($node$, $m\_out$) \\
    \textbf{11 \quad}  \\
    \textbf{12 for} each $node$ in $graph$ do:\\
    \textbf{13 \quad} $in\_edges$ $\leftarrow$ $graph[node].incoming$ \\
    \textbf{14 \quad} $out\_edges$ $\leftarrow$ $graph[node].outgoing$ \\
    \textbf{15 \quad} $m\_in$ $\leftarrow$ CreateMinHash($in\_edges$) \\
    \textbf{16 \quad} $m\_out$ $\leftarrow$ CreateMinHash($out\_edges$) \\
    \textbf{17 \quad} $candidates\_in$ $\leftarrow$ LSH.query($m\_in$) \\
    \textbf{18 \quad} $candidates\_out$ $\leftarrow$ LSH.query($m\_out$) \\
    \textbf{19 \qquad} $silmilar\_nodes$ $\leftarrow$ Union($candidates\_in$, $candidates\_out$)\\
    \textbf{20 \quad}$overlap\_ratio$ $\leftarrow$ $\vert$ $silmilar\_nodes$ $\vert$ / $\vert$ $graph.nodes$ $\vert$ \\
    \textbf{21 \quad}$overlap\_ratios$[$node$] = $overlap\_ratio$\\
    \textbf{22 } OutputToFile ($overlap\_ratios$)\\

    \bottomrule
\end{tabular}
\end{table}

\section{EXPERIMENTS}
In this section, an overview of the experimental setup is introduced, and the proposed method is evaluated based on the experimental results.

\subsection{Datasets}
We select four real directed attribute graph datasets from different fields for experiments, including \textit{CollegeMsg}, \textit{Email}, \textit{Google+} and \textit{Twitter}, which can be available at http://snap.stanford.edu/data. \autoref{tab:2} gives the details of the datasets. 

\begin{table}[hb]
  \caption{Detailed information of datasets}
  \label{tab:2}
  \begin{tabular}{ccccl}
    \toprule
    Dataset & Nodes & Edges & Domains &Sources \\
    \midrule
    \textit{CollegeMsg} & 1,899 & 59,835 & Temporal network & ~\cite{panzarasa2009patterns} \\
    \textit{Email} & 1,005 & 25,571 & Email-Eu-core network & ~\cite{yin2017local, leskovec2007graph}\\
    \textit{Google+} & 1,651 & 166,292 & Social network &~\cite{leskovec2012learning}\\
    \textit{Twitter} & 475 & 13,289 & Interaction network &~\cite{fink2023centrality}\\

  \bottomrule
\end{tabular}
\end{table}

\subsection{Experimental Settings}
In this part, we designed three groups of experiments to evaluate the effect and efficiency of node pricing based on substitutability with dominator tree.

\textbf{A. Effectiveness of pricing based on substitutability:} We evaluate the effectiveness of pricing methods based on substitutability by analyzing the outcomes of various pricing strategies. Specifically, we compare the method proposed in this paper with two well-established pricing methods: the Shapley value and Information Entropy, both of which are grounded in different theoretical frameworks.

From the perspective of destructiveness, the Shapley value is a prominent solution concept in cooperative game theory that provides a fair distribution of payoffs among players based on their contributions ~\cite{WINTER20022025}. We employ this method because it allows for a nuanced analysis of how individual nodes in a graph contribute to overall value creation. The Shapley value calculates the marginal contributions generated by each node by considering the contributions of subsets of nodes, thereby pricing each node within the graph. Due to the high complexity of computing the Shapley value, which is an NP-hard problem, we introduce Monte Carlo Shapley value to price the nodes.

From the perspective of scarcity, Information Entropy serves as a measure of uncertainty within a dataset. The greater the Information Entropy, the less uncertainty there is in the occurrence of an event, and the higher the probability of correctly estimating it. Therefore, the greater the Information Entropy, the more effective the information content is, and the higher the transaction price. This method fully considers the scarcity of data assets and focuses more on the effective quantity and distribution of data compared to its content and quality ~\cite{Li2017}. 
In graph data, the quantity and distribution of nodes can also be analyzed based on their structural and attribute information, in order to price each node.

It is important to note that the assumptions underlying these methods differ significantly, the method proposed in this paper and the Information Entropy assumption that the price of the entire graph is known for pricing, while Shapley calculates the marginal revenue of nodes by assuming the subset price in the graph. In practical applications, relying on pricing methods that assume known node prices is insufficiently accurate.

\textbf{B. Effectiveness of dominator tree:} This part aims to demonstrate the effectiveness of using the dominator tree to extract structural information. A comparison is made between the dominator tree and other methods based on structural information. Specifically, two different methods that consider structural information are introduced, Degree Centrality and Eigenvector Centrality. These two methods respectively take into account the local and global structural information of nodes. 

Degree Centrality ~\cite{Zhang201703, 7065911} is a simple yet effective measure that reflects the local connectivity of each node. By considering the degree of a node, we can obtain an indication of its importance within its immediate neighborhood. This is particularly useful in situations where local interactions play a significant role. On the other hand, Eigenvector Centrality ~\cite{BONACICH2007555} considers the global structure of the graph. It assigns higher importance to nodes that are connected to other important nodes. This method provides a more comprehensive view of the graph structure and helps to identify nodes that are crucial for the overall functioning of the graph data.

This part demonstrates the effectiveness of extracting relationships between nodes based on dominator trees by comparing different methods for extracting graph structures.


\textbf{C. Efficiency and scalability of our method:} We evaluate the running time of these sets of experiments to analyse the efficiency of our method. Furthermore, to assess the scalability of the approach presented in this paper, we also randomly constructed ten datasets of varying sizes.

\subsection{Experimental Results}

We choose four real datasets for our study: \textit{CollegeMsg}, \textit{Email}, \textit{Google+} and \textit{Twitter}, each of which has the different density, with the first two datasets having sparser edge distributions and the last two datasets having denser ones. The primary difference between the graphs lies in their structural information. Dense graphs possess a more extensive structural information, which facilitates the flow of information between nodes, promoting greater sharing of information among them. As a result, it is likely that the pricing differences between nodes in dense graphs are comparatively smaller compared to those in sparse graphs.

\begin{figure*}
  \centering
  \includegraphics[width=\linewidth]{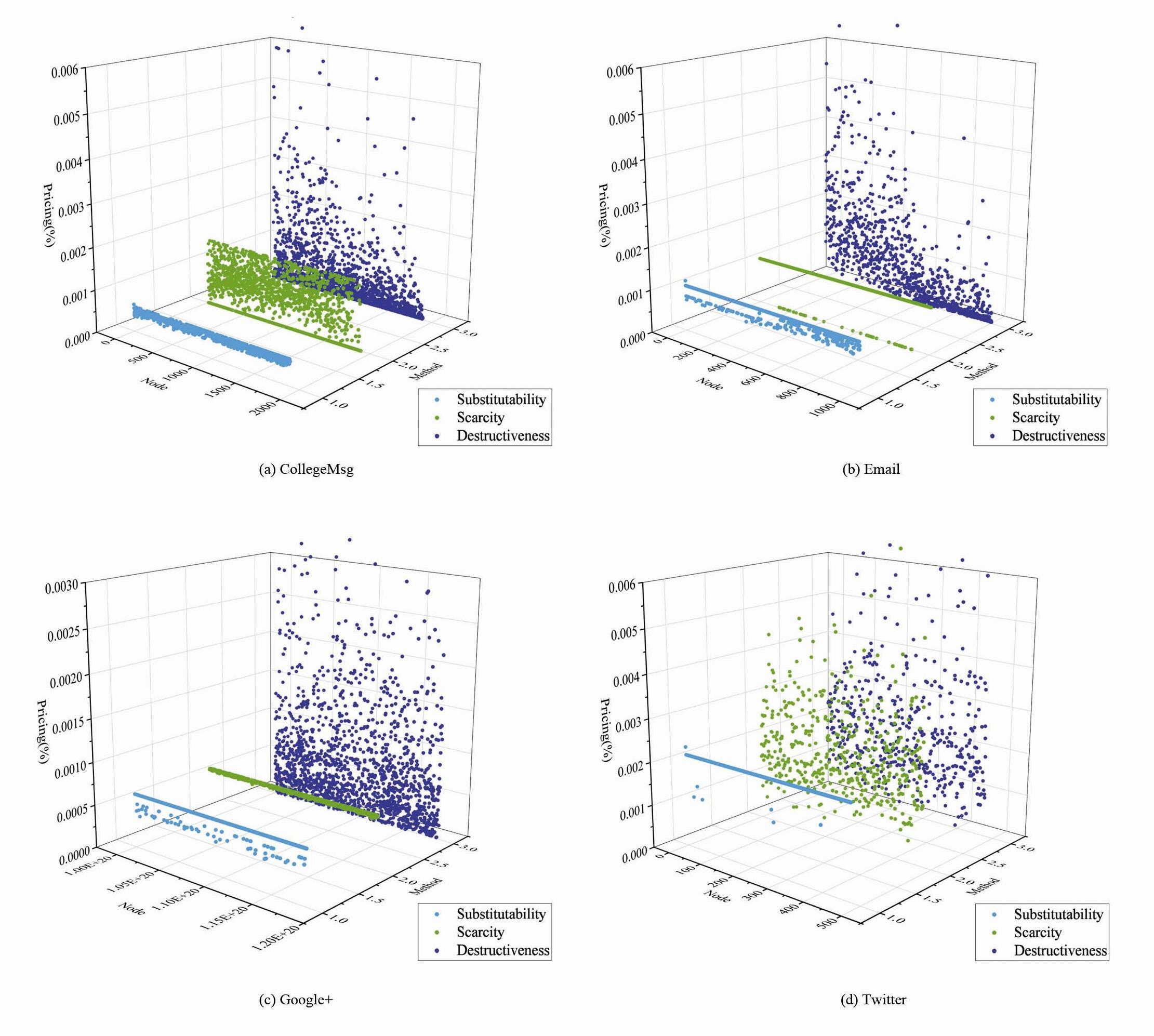}
  \caption{Exp.A: Effects of pricing based on substitutability.}
  \label{fig:s1}
\end{figure*}

\textbf{A. Effectiveness of pricing based on substitutability:} \autoref{fig:s1} shows the pricing effects calculated from substitutability, destructiveness, and scarcity. Our method mainly targets directed connected graphs, so some preprocessing was performed on the dataset before computation, and isolated nodes were removed. As shown in \autoref{fig:s1}, it illustrates the results obtained through pricing methods based on substitutability for graph data with varying edge densities. It is apparent that the results generally show a trend where nodes are predominantly clustered within a certain range, with some nodes exhibiting outlier distributions. However, there are notable distinctions in their distribution patterns. When pricing graph data with relatively sparse edge densities, the distribution becomes more scattered, implying that the differences between nodes are more pronounced. In contrast, when pricing graph data with denser edge densities, the distribution is more concentrated. This pricing scenario aligns well with the characteristics of different graph data types. In sparse graphs, the exchange of information between nodes is limited, leading to significant differences among nodes. In contrast, in dense graphs, the frequency of information exchange between nodes is higher, thereby mitigating the differences in information among nodes.

Secondly, for pricing method based on destructiveness, this calculation method involves a large number of subset permutation combinations and probability calculations, which are very sensitive to subtle changes in data. Under different datasets or scenarios, due to the large number and significant differences in possible alliance combinations, the fluctuation range of the final pricing results will be large, showing higher dispersion compared to the other two methods.

And for scarcity, which primarily takes into account the distribution of attribute information, given the relatively even distribution of information in the dataset, the pricing results tend to be average. However, this average pricing fails to adequately reflect the information differences between nodes. It is important to note that the dataset (a) is represented in the form of timestamps, which are discrete labels; the attribute information in datasets (b) and (c) is categorical; while dataset (d) contains actual numerical weights. This distinction results in a relatively uniform pricing structure based on scarcity for the first three datasets, whereas dataset (d) exhibits greater dispersion in its pricing outcomes. This observation highlights a limitation of scarcity-based pricing, as it may not adequately account for the nuances associated with different types of attribute information. The reliance on categorical attributes can lead to oversimplified pricing models, which may fail to capture the complexities inherent in datasets characterized by continuous numerical values. For dataset (a), although it exhibits a relatively dispersed pricing distribution, the issue lies in the fact that it assigns a price of zero to certain nodes, which is unrealistic in actual trading scenarios. Such limitations emphasize the need for more sophisticated approaches that can effectively integrate and analyze diverse types of attribute information to enhance pricing accuracy. Our method aims to address this limitation and provide a more nuanced pricing approach that takes into account the unique characteristics of each node.

Experiment A demonstrates that when pricing sparse graphs based on substitutability, due to the limited structural information in the graph, distinct pricing results can be obtained. In contrast, for dense graphs, with their rich structural information, nodes can exchange information, leading to convergent pricing results. In contrast, the pricing methods based on destructiveness and scarcity fail to adequately account for the structural information of the graph.

\begin{figure*}
  \centering
  \includegraphics[width=\linewidth]{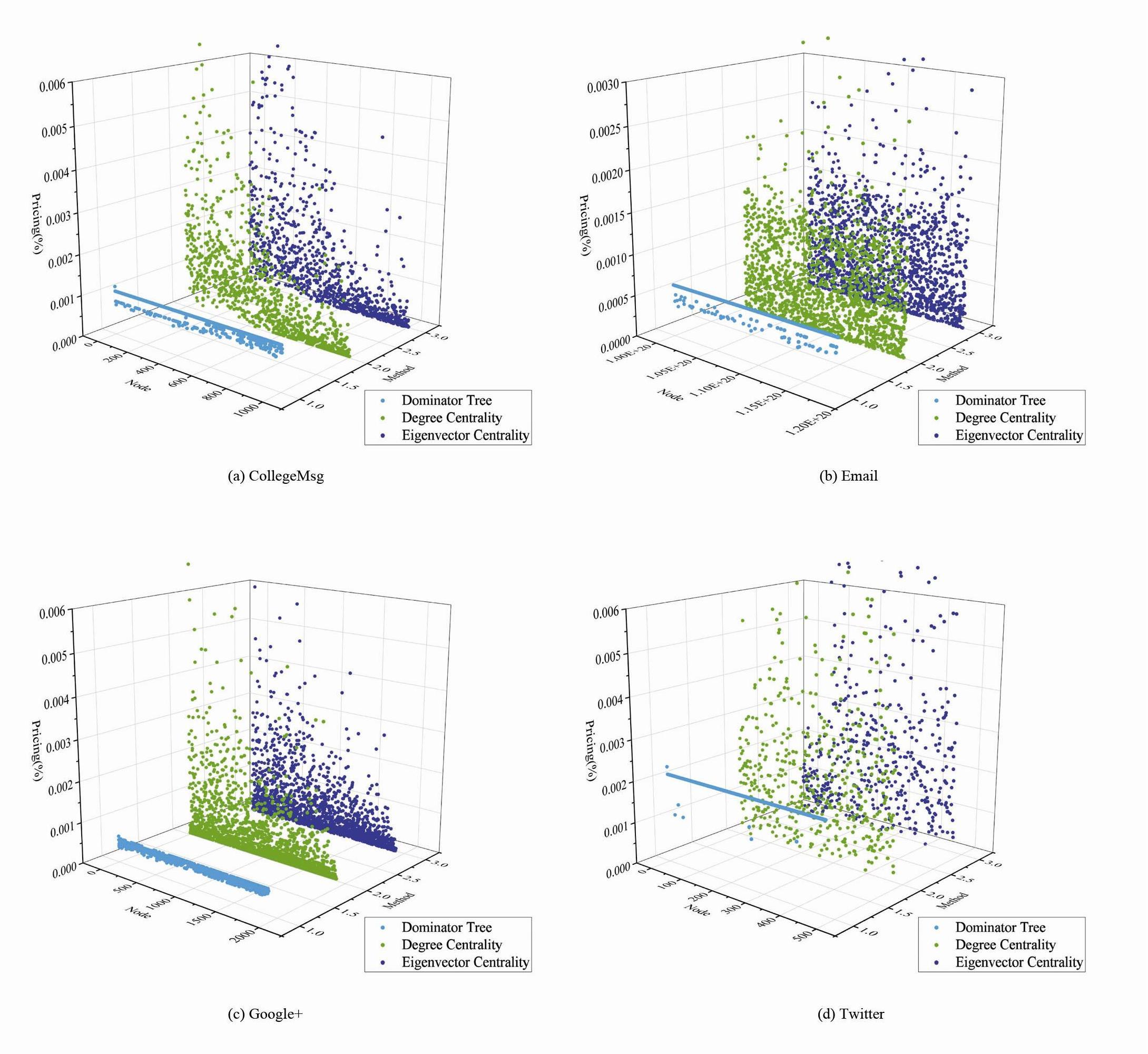}
  \caption{Exp.B: Effects of dominator tree}
  \label{fig:1229s2}
\end{figure*}

\textbf{B. Effectiveness of dominator tree:} As shown in \autoref{fig:1229s2}, dominator tree, degree centrality, and eigenvector centrality all consider the connectivity and influence of nodes to a certain extent. The dominator tree reflects the control ability of nodes through path dependency. Degree centrality directly considers the degree of nodes, that is, the number of connections. Eigenvector centrality comprehensively considers the importance of nodes themselves and the importance of nodes connected to them. Although these methods start from different angles, in dense graphs, due to the close connection between nodes, they can all capture the important features of nodes in the graph, resulting in relatively similar node pricing results. In addition, because the dominator tree mainly focuses on path dependency and the nodes in dense graphs have many possible paths, leading to a more concentrated pricing result with less differentiation. It is noteworthy that the pricing results of the other two methods also yield instances where prices are set to zero, rendering them unsuitable for real-world trading applications. Furthermore, in the context of dense graphs, acquiring information from certain nodes allows for the inference of additional insights based on structural data. The calculation methods for degree centrality and eigenvector centrality can lead to significant price disparities among nodes, thereby facilitating the emergence of arbitrage opportunities ~\cite{ross1976arbitrage}. Therefore, it can be seen that the dominator tree is a reasonable and effective method for obtaining structural information.

\textbf{C. Efficiency and scalability of our method:} Evaluating the efficiency and scalability of our proposed method is crucial for its practical application in real-world scenarios. To comprehensively assess these aspects, we measured the running time of our method on the four real-world datasets: $CollegeMsg$, $Email$, $Google+$, and $Twitter$.

\begin{figure}
  \centering
  \includegraphics[width=\linewidth]{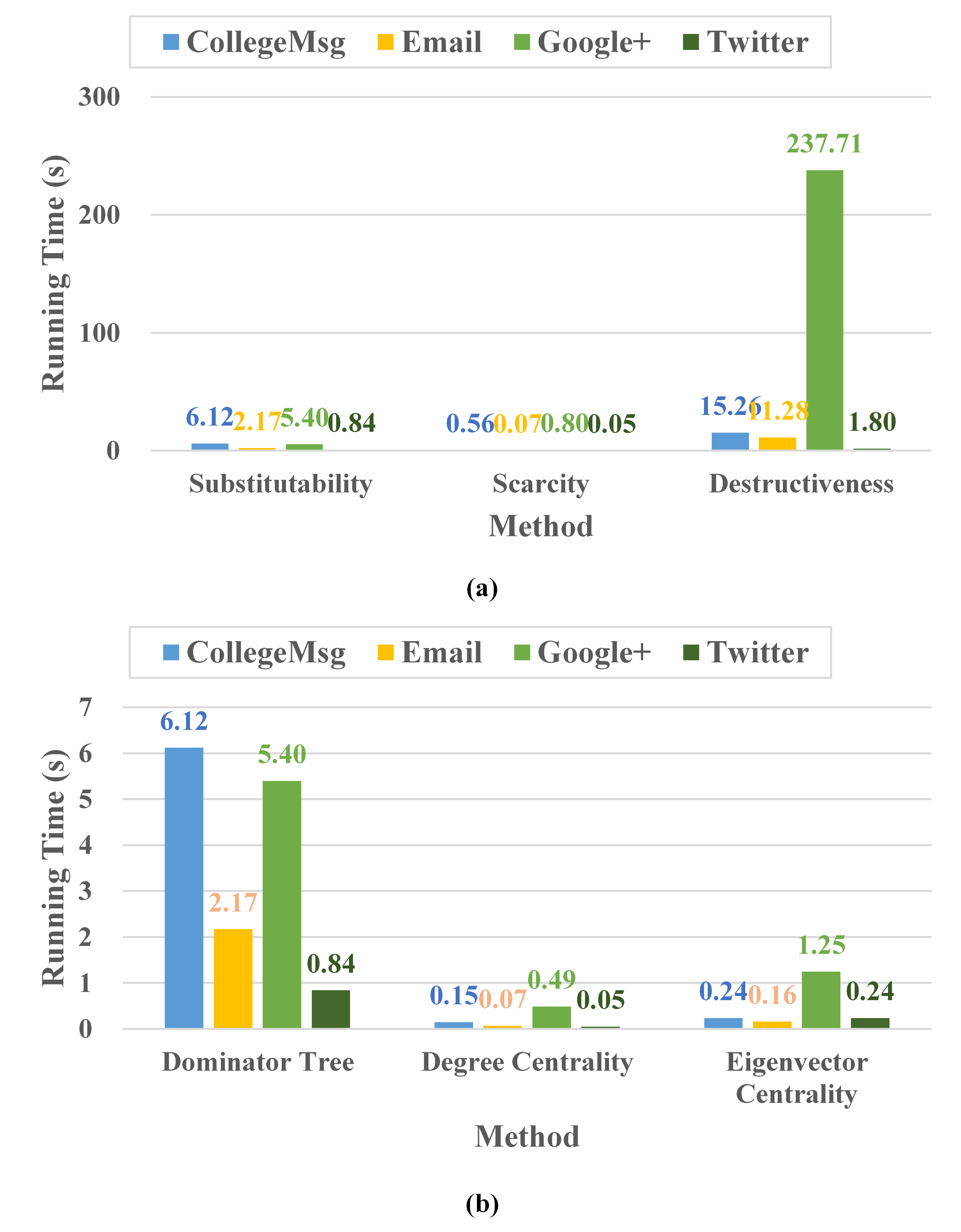}
  \caption{Running time of different methods.}
  \label{fig:s3}
\end{figure}

Our method's running times on these datasets are 6.12s, 2.17s, 5.40s and 0.84s respectively as shown in \autoref{fig:s3}. When compared with the scarcity-based method, which is relatively simple, our method has a slightly longer running time. This is mainly because our approach is designed to handle the complex structural information of graph data, incorporating the dominator tree and other advanced concepts. In contrast, the scarcity - based method often focuses solely on attribute information, which simplifies its computational process. However, when compared with the destructiveness - based method, our method is more efficient. The destructiveness - based method, such as calculating the Shapley value, involves a large number of subset permutation combinations and probability calculations. For example, in a graph with $n$ nodes, calculating the exact Shapley value has a time complexity of $O(2^n)$, which makes it computationally expensive. Our method, on the other hand, with a time complexity of $O(mlogn)$ for constructing the dominator tree using the Lengauer-Tarjan algorithm and $O(n)$ for the MinHash-LSH approximation in large - scale graphs, significantly reduces the computational burden.

To further demonstrate the scalability of our method, we compared the original algorithm with the improved version using the MinHash-LSH approximation. In large-scale graphs, the structural similarity analysis of the original algorithm has an $O(n^2)$ computational complexity when using exact pairwise comparisons. This quadratic growth makes precise calculations computationally prohibitive and memory-intensive, especially in real - world network datasets with a large number of nodes (e.g., social networks with \(>10^6\) nodes).

\begin{figure}
  \centering
  \includegraphics[width=\linewidth]{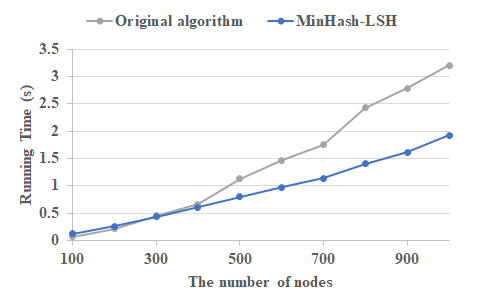}
  \caption{Running time with different scales.}
  \label{fig:timee}
\end{figure}

The MinHash-LSH approximation in our improved method addresses these limitations. By encoding structural roles via edge-set fingerprints rather than exhaustive comparisons and using locality-sensitive hashing for sublinear-time similarity searches, the improved method achieves a \(O(n)\) time complexity. As shown in \autoref{fig:timee}, as the scale of data increases and the structural information to be processed grows, the running time of our improved method essentially follows a linear trend. This indicates that our method can handle larger datasets more efficiently compared to the original algorithm. For instance, when the number of nodes in the dataset doubles, the running time of the original algorithm would increase four-fold due to its quadratic complexity, while the running time of our improved method would only double, showing its superiority in scalability.

In summary, our method strikes a balance between handling complex graph structures and computational efficiency. Although it may have a slightly longer running time compared to the simplest methods in some cases, it outperforms more complex methods like those based on destructiveness. Moreover, the use of the MinHash-LSH approximation in our improved method significantly enhances its scalability, making it suitable for large - scale graph data pricing tasks. 

\section{RELATED WORK}
The pricing of graph data has emerged as a critical challenge in recent years, driven by the increasing complexity of graph-structured data and its diverse applications in social networks, recommendation systems, and knowledge graphs ~\cite{muschalle2013pricing, birnbaum2007relational, choudhary2010use, yu2017data, sen2013survey}. Early studies in data pricing focused primarily on relational databases, where pricing models such as cost-based pricing and value-based pricing were developed to address query evaluation costs and user utility ~\cite{9300226, 10.11452770870}. However, the structural characteristics of graph data, such as node interdependencies and topological complexity, necessitate specialized pricing approaches that account for graph-specific properties.

Subsequent research shifted focus to graph query pricing, with notable contributions including the GQP framework by Chen et al.~\cite{chen2022gqp}, which introduced dynamic pricing based on graph traversal costs. This work addressed the computational efficiency of pricing by reusing precomputed price points, yet it retained the assumption of fixed node prices. Hou et al.~\cite{hou2023scalable} extended this line of inquiry by developing pricing functions for incomplete graph queries, leveraging subgraph matching techniques to derive query costs. These approaches laid the groundwork for query-centric pricing but remained agnostic to the intrinsic value of individual nodes within the graph structure.

A second research stream explored influence-based pricing in social networks, where node values are determined by their propagation potential. Zhu et al.~\cite{zhupricing} proposed a method based on expected impact propagation, enabling advertisers to select optimal marketing initiators without full network knowledge. However, this work relied on simplified propagation models and did not account for dynamic network evolutions. More recently, Sun et al.~\cite{SUN2024122815} advanced this area using graph autoencoders to predict node influence through augmented neighborhood subgraphs, incorporating Shapley values to balance pricing fairness and utility. Despite these innovations, their framework remained dependent on complete cascade data for training, limiting its applicability in real-world scenarios with incomplete propagation histories.

Complementary efforts focused on privacy-preserving pricing mechanisms for graph statistics. Chen et al.'s GSHOP~\cite{10597944} introduced differential privacy to query pricing, balancing affordability and arbitrage prevention through noisy answer perturbations. While GSHOP enhanced fairness in query-level pricing, it did not explicitly model node-value interdependencies arising from graph structure.

Notwithstanding these advances, existing approaches share two fundamental limitations: (1) they either assume predefined node prices or derive prices from isolated node attributes, neglecting the structural roles nodes play in graph dynamics; and (2) they rely on historical data for calculations and fail to price static graphs of newly entered markets.. To bridge these gaps, our work introduces a novel pricing framework grounded in economic substitutability principles, leveraging dominance trees to model node structural influence and derive rational, interdependent node prices. This approach represents a paradigm shift from query-centric pricing to a graph-wide valuation method that accounts for both topological properties and economic substitutability.

Therefore, this paper grounded in the economic principle of substitutability, introduces dominance trees to acquire structural information about nodes, thereby providing a rational pricing strategy for them.

\section{CONCLUSION}
In this paper, we focus the problem of node pricing based on substitutability and complementarity. We adopt the dominator tree with Lengauer-Tarjan algorithm to extract the dominance relationships. Extensive experiments on real or random graph datasets demonstrate the effectiveness and efficiency of our pricing method. In the future, we will study pricing methods for other types of data, such as pricing relational data based on substitutability, and even for any other type of data commodity.

\begin{acks}
 This work was supported by the China National Social Science Fund Post-Funding Project “Research on Data Trading Issues”(22FJYB022) and the Innovation and Talent Base for Digital Technology and Finance Project (B21038).
\end{acks}


\bibliographystyle{ACM-Reference-Format}
\bibliography{main}

\end{document}